\documentclass[twocolumn,superscriptaddress,amsmath, amssymb, amsfonts,preprintnumbers,aps,prd,longbibliography,
nofootinbib
]{revtex4-1}

\renewcommand{\sec}[1]{\textbf{#1\,--}}

\usepackage{graphicx}
\usepackage{dcolumn}
\usepackage{bm}
 \usepackage{amstext}
 \usepackage{amssymb}
 \usepackage{amsmath}
 \usepackage{graphicx}
 \usepackage{color}
 \usepackage{bbold}
 \usepackage{delimset} 
\usepackage[caption=false,justification=justified]{subfig}
\usepackage[colorlinks=true]{hyperref}
\usepackage{orcidlink}
\usepackage{float}
\usepackage{soul}

\newcommand{\gusNotes}[1]{}

\DeclareFontFamily{OT1}{pzc}{}
\DeclareFontShape{OT1}{pzc}{m}{it}{<-> s * [1.350] pzcmi7t}{}
\DeclareMathAlphabet{\mathpzc}{OT1}{pzc}{m}{it}

\def\nn{\nonumber}

\newcommand{\be}{\begin{equation}}
\newcommand{\ee}{\end{equation}}
\newcommand{\ba}{\begin{align}}
\newcommand{\ea}{\end{align}}
\ifx\genfrac\sdflkaj\else\fi

\begin{document}

\preprint{HU-EP-24/16-RTG}

\title{Post-Minkowskian Theory Meets the Spinning \\ Effective-One-Body Approach for Bound-Orbit Waveforms}

\author{Alessandra Buonanno\,\orcidlink{0000-0002-5433-1409}}
\affiliation{Max Planck Institute for Gravitational Physics (Albert Einstein Institute), Am M\"uhlenberg 1, 14476 Potsdam, Germany}
\affiliation{Department of Physics, University of Maryland, College Park, MD 20742, USA}

\author{Gustav Mogull\,\orcidlink{0000-0003-3070-5717}}
\affiliation{Max Planck Institute for Gravitational Physics (Albert Einstein Institute), Am M\"uhlenberg 1, 14476 Potsdam, Germany}
\affiliation{%
  Institut f\"ur Physik und IRIS Adlershof, Humboldt Universit\"at zu Berlin,
  Zum Gro{\ss}en Windkanal 2, 12489 Berlin, Germany
}

\author{Raj Patil\,\orcidlink{0000-0002-7055-0345}}
\affiliation{Max Planck Institute for Gravitational Physics (Albert Einstein Institute), Am M\"uhlenberg 1, 14476 Potsdam, Germany}
\affiliation{%
  Institut f\"ur Physik und IRIS Adlershof, Humboldt Universit\"at zu Berlin,
  Zum Gro{\ss}en Windkanal 2, 12489 Berlin, Germany
}

\author{Lorenzo Pompili\,\orcidlink{0000-0002-0710-6778}}
\affiliation{Max Planck Institute for Gravitational Physics (Albert Einstein Institute), Am M\"uhlenberg 1, 14476 Potsdam, Germany}

\begin{abstract}
  Driven by advances in scattering amplitudes and worldline-based
  methods, recent years have seen significant progress in our
  ability to calculate gravitational two-body scattering
  observables. These observables effectively encapsulate the
  gravitational two-body problem in the weak-field and high-velocity
  regime (post-Minkowskian, PM), with applications to the bound two-body
  problem and gravitational-wave modeling. We 
  leverage PM data to construct a complete
  inspiral-merger-ringdown waveform model for non-precessing spinning
  black holes within the effective-one-body (EOB) formalism: \texttt{SEOBNR-PM}. 
This model is closely based on the highly
  successful \texttt{SEOBNRv5} model, used by the LIGO-Virgo-KAGRA Collaboration, with its key new feature being an
  EOB Hamiltonian derived by matching the two-body
  scattering angle in a perturbative PM expansion. The
  model performs remarkably well, showing a median mismatch against
  441 numerical-relativity (NR) simulations that is somewhat lower
  than a similarly calibrated version of \texttt{SEOBNRv5}. Comparisons of the
  binding energy with NR also demonstrate better agreement than
  \texttt{SEOBNRv5}, despite the latter containing additional calibration to NR
  simulations.
\end{abstract}

\maketitle

\sec{Introduction} 
Since the initial detection of a gravitational wave (GW) from a
  binary-black-hole (BBH) merger~\cite{LIGOScientific:2016aoc}, the
  LIGO-Virgo-KAGRA
  Collaboration~\cite{LIGOScientific:2014pky,VIRGO:2014yos,KAGRA:2020tym}
  and independent analyses have identified about 100 mergers of
  compact binaries~\cite{LIGOScientific:2017vwq,KAGRA:2021vkt,Nitz:2021zwj,Olsen:2022pin,Mehta:2023zlk,Wadekar:2023gea}. These
  observations have begun to reveal the distributions of BH masses and spins~\cite{KAGRA:2021duu}, improved constraints on the
  neutron-star equation of state~\cite{LIGOScientific:2018cki},
  obtained independent measurements of the Hubble-Lema\^itre
  parameter~\cite{LIGOScientific:2017adf, LIGOScientific:2021aug}, and
  validated General Relativity~\cite{LIGOScientific:2016lio,LIGOScientific:2020tif,LIGOScientific:2021sio}.

Enhancements in the sensitivity of current GW detectors, coupled
  with the development of next-generation observatories like the
  Einstein Telescope and Cosmic
  Explorer~\cite{Punturo:2010zz,Kalogera:2021bya,Reitze:2019iox}, as
  well as future space-based detectors such as
  LISA~\cite{LISA:2017pwj}, TianQin~\cite{TianQin:2015yph} or
  Taiji~\cite{TaijiScientific:2021qgx}, are poised to dramatically
  increase the number of detectable GW sources. These advancements
  will enable observations with a signal-to-noise ratio  up to
  two orders of magnitude higher than what is currently
  achievable~\cite{Borhanian:2022czq}, necessitating a commensurate
  improvement in the accuracy of waveform models. Recent
  research~\cite{Dhani:2024jja} has demonstrated that even
  state-of-the-art waveform models, designed for quasi-circular,
  spin-precessing BBHs, exhibit systematic biases
  when applied to future LIGO-Virgo-KAGRA runs and next-generation detectors. This bias becomes
  pronounced, especially for high-spin rates and
  significant asymmetries in spins and masses. Addressing the
  challenge of waveform accuracy is essential to realizing the full
  scientific potential of future runs and 
  detectors~\cite{KAGRA:2013rdx,Evans:2021gyd,Bogdanov:2022faf,Borhanian:2022czq,Colpi:2024xhw}
and avoiding false claims of General Relativity violations~\cite{Toubiana:2023cwr,Gupta:2024gun}. 

Waveform models for compact binaries are crafted by
  synergistically combining analytical and numerical 
  relativity (NR) results. NR tackles the
  formidable task of solving Einstein's equations on
  supercomputers~\cite{Pretorius:2005gq,Campanelli:2005dd,Baker:2005vv},
  a process notorious for its time-intensive nature. On the other
  hand, perturbative methods are used to obtain approximate solutions to
  Einstein's equations, offering analytic formulas that are swift to evaluate. 
Three primary perturbative approaches have been
  developed: post-Newtonian (PN)
  theory~\cite{Futamase:2007zz,Blanchet:2013haa,Porto:2016pyg,Schafer:2018kuf,Levi:2018nxp,
  	Jaranowski:1997ky,
  	Damour:2014jta,Jaranowski:2015lha,Bernard:2015njp,Bernard:2016wrg,Damour:2016abl,
  	Blanchet:2023sbv,Blanchet:2023bwj,
  Foffa:2019rdf,Foffa:2019yfl,Blumlein:2020pog}
  applicable in the weak-field and small-velocity limit,
  post-Minkowskian (PM)
  theory~\cite{Westpfahl:1979gu,Westpfahl:1980mk,Bel:1981be,Westpfahl:1985tsl,schafer1986adm,Ledvinka:2008tk,Buonanno:2022pgc,Travaglini:2022uwo,Bjerrum-Bohr:2022blt,Kosower:2022yvp}
  in the weak-field regime, and the gravitational-self force (GSF)
  formalism~\cite{Mino:1996nk,Quinn:1996am,Barack:2001gx,Barack:2002mh,Gralla:2008fg,Detweiler:2008ft,Keidl:2010pm,vandeMeent:2017bcc,Pound:2012nt,Pound:2019lzj,Gralla:2021qaf,Pound:2021qin,Warburton:2021kwk,Wardell:2021fyy}
  for the small mass-ratio limit. These analytical results are then
  synthesized in the effective-one-body (EOB) approach~\cite{Buonanno:1998gg,Buonanno:2000ef,Damour:2000we,Damour:2001tu,Buonanno:2005xu},
  which efficiently resums the perturbative calculations for the
  inspiral while retaining known non-perturbative results for
  BHs, achieving high accuracy for current observing
  runs via calibration to NR~\cite{Buonanno:2006ui,Buonanno:2007pf,Damour:2007yf,Pan:2011gk,Damour:2012ky,Pan:2013rra,Taracchini:2013rva,Bohe:2016gbl,Babak:2016tgq,Nagar:2018zoe,Ossokine:2020kjp,Gamba:2021ydi,Nagar:2023zxh,Pompili:2023tna,Ramos-Buades:2023ehm}.

  Thus far, the EOB waveform models utilized by the LIGO-Virgo-KAGRA
  Collaboration\footnote{The LIGO-Virgo-KAGRA Collaboration also utilizes
    phenomenological frequency- and time-domain waveform models, which
    are built by combining PN, EOB, and NR
    results~\cite{Ajith:2007qp,Pratten:2020ceb,Estelles:2021gvs}. Additionally,
    it employs waveforms directly interpolated from NR
    simulations~\cite{Blackman:2017pcm,Varma:2019csw}, where
    available.} have primarily relied on resummations of the PN
  expansion, with the exception of
  Refs.~\cite{Ramos-Buades:2023ehm,vandeMeent:2023ols}, which included
  second-order GSF results~\cite{Warburton:2021kwk} for the
  gravitational modes and radiation-reaction force. Given recent
  advancements in PM~\cite{Damour:2016gwp,
  		Cheung:2018wkq,
  		Bern:2019nnu,Bern:2022kto,Bern:2024adl,
  		Kosower:2018adc,
  		Bjerrum-Bohr:2018xdl,
  		Cristofoli:2019neg,
  		Damgaard:2019lfh,
  		Brandhuber:2021eyq,
  		Vines:2017hyw,
  		Kalin:2020mvi,Kalin:2020fhe,
  		Mogull:2020sak,Jakobsen:2021zvh,Jakobsen:2022psy,Bhattacharyya:2024aeq}
  and GSF~\cite{Warburton:2021kwk,Wardell:2021fyy}, there is
  now significant interest in exploring and developing waveform models
  that combine information from various perturbative methods in
  innovative ways. The aim is to address the waveform-accuracy
  challenge. In this regard, the PM approach is particularly
  interesting since an (n + 1) PM-order ($G^{n+1}$) Hamiltonian includes all information up to
  the nPN order (wherein $v^2/c^2\sim GM/(rc^2)$), and additional weak-field, high-velocity information
  from infinitely higher PN orders, making it suitable for systems
  with high velocities or large eccentricities at fixed periastron
  distances~\cite{Khalil:2022ylj}.
  Using sophisticated quantum-field-theory-based methods,
  tremendous progress has been made on the precision PM frontier using both
  scattering amplitudes~\cite{Bern:2021dqo,Bern:2021yeh,Damgaard:2023ttc} and
  worldlines~\cite{Dlapa:2021vgp,Dlapa:2023hsl,Jakobsen:2023ndj,Jakobsen:2023pvx}.
  This  progress is largely due to a  blend of a clever and efficient organization of perturbative calculations
  and formal mathematical developments in understanding the properties
  of multi-loop integrals \cite{Parra-Martinez:2020dzs,Weinzierl:2022eaz,Abreu:2022mfk,Blumlein:2022qci,Frellesvig:2023bbf,Frellesvig:2024zph,Klemm:2024wtd}.
These developments were primarily driven over the past several decades to address
  precision-collider physics. Furthermore, several sophisticated
  techniques, such as generalized unitarity \cite{Bern:1994cg,Bern:2011qt},
  double copy~\cite{Bern:2010ue,Bern:2019prr,Bern:2022wqg,Adamo:2022dcm},
  supersymmetry~\cite{Jakobsen:2021zvh} and
  massive higher spins \cite{Chung:2018kqs,Arkani-Hamed:2019ymq,Bern:2020buy,Bautista:2021wfy,Bautista:2022wjf,Chiodaroli:2021eug,Aoude:2023vdk,Cangemi:2023ysz}
 have also been used to further enhance these computations.

\begin{figure}[t]
	\vspace{-8pt}
	\includegraphics[width=.46\textwidth]{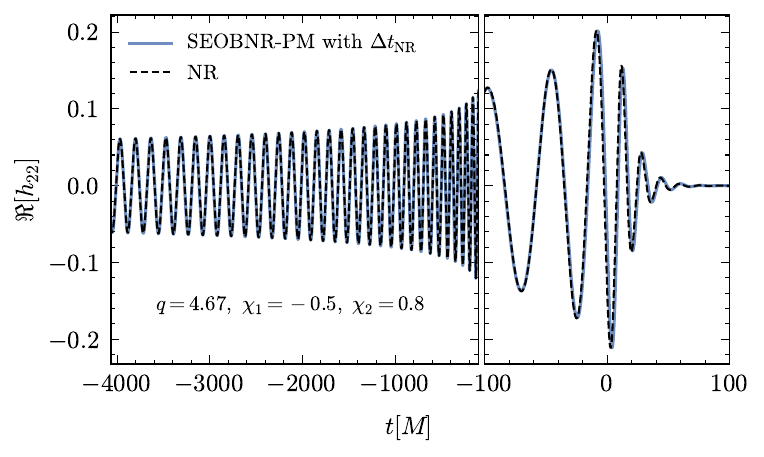}
	\captionsetup{skip=5pt}
	\caption{An \texttt{SEOBNR-PM} inspiral-merger-ringdown
		waveform generated using the \texttt{pySEOBNR} code~\cite{Mihaylov:2023bkc}, 
		compared against the NR simulation \texttt{SXS:BBH:1445}, 
		after a low-frequency alignment by a time and phase shift. 
		The time $t = 0$ corresponds to the peak of the $(2,2)$ mode of the NR waveform.
		\label{fig:khal}
	}
	\vspace{-8pt}
\end{figure}

In this Letter, leveraging on Refs.~\cite{Khalil:2023kep,Pompili:2023tna,Buonanno:2024vkx}, we present the first PM-informed spinning EOB waveform model: \texttt{SEOBNR-PM}, encompassing the inspiral, as well as the merger and ringdown phases.
This model incorporates the most recent findings from PM theory into the EOB Hamiltonian,
and it is mildly calibrated to NR waveforms. The \texttt{SEOB-PM} Hamiltonian (so named as it does not include an NR calibration term) 
includes the non-spinning (conservative) 4PM~\cite{Bern:2021yeh,Dlapa:2021vgp} and spinning 5PM terms~\cite{Guevara:2018wpp,Bern:2020buy,Kosmopoulos:2021zoq,Chen:2021kxt,Jakobsen:2022fcj,Jakobsen:2022zsx,FebresCordero:2022jts,Jakobsen:2023ndj}, alongside the known non-spinning 4PN \cite{Damour:2014jta,Jaranowski:2015lha,Bernard:2015njp,Bernard:2016wrg,Damour:2016abl} contributions, which also corrects the tails from unbound to bound orbits up to that order.\footnote{We plan to update  
our \texttt{SEOBNR-PM} model in the near future to include the recent findings in Ref.~\cite{Dlapa:2024cje}, which enables derivation of the local-in-time 4PM EOB Hamiltonian for bound orbits.} Our PM counting is a physical one, with spin orders contributing in addition to loop orders (see Table II in Ref.~\cite{Buonanno:2024vkx}).  We construct our \texttt{SEOBNR-PM} model within the \texttt{pySEOBNR} code~\cite{Mihaylov:2023bkc}, which was recently built to make 
the development of \texttt{SEOBNR} models, including  the calibration to NR waveforms, highly efficient. As an example, we show in Fig.~\ref{fig:khal} the agreement between the (mildly) calibrated 
\texttt{SEOBNR-PM} and NR for a spinning BBH coalescence.

\sec{EOB framework for waveforms} We use geometric units $G = 1 = c$, and set $M = m_1 +m_2$ and $\nu = m_1 m_2/M^2$, 
where $m_1$ and $m_2$ are the BH's masses. In the EOB formalism, the binary's 
conservative dynamics is described by the EOB Hamiltonian
$H_{\rm EOB} = M\, \sqrt{1 + 2 \nu (H_{\rm eff}/\mu -1) }$,
where $H_{\rm eff}$ is the Hamiltonian of an
effective test body of mass $\mu = \nu \,M$ moving in the (deformed) Kerr spacetime, with $0 \leq \nu \leq 1/4$ the deformation parameter. We also introduce the mass ratio $q = m_1/m_2 > 1$. 
We limit to nonprecessing spins (i.e., aligned spins) and introduce the spin lengths $a_i=m_i\chi_i$, with $a_\pm=M\chi_\pm=a_1\pm a_2$.

In the center-of-mass frame, the inspiral-plunge dynamics, for aligned-spin BHs, is computed from the EOB equations of motion \cite{Buonanno:2000ef,Pan:2011gk}:
\begin{subequations}
\begin{align}\label{eq:dyn}
  \dot{r}&=\frac{\partial H_{\rm EOB}}{\partial p_r}\,,  & \dot{p}_r &= -\frac{\partial H_{\rm EOB}}{\partial r} + \frac{p_r}{p_\phi} \mathcal{F}_\phi\,,\\
  \dot{\phi}&=\frac{\partial H_{\rm EOB}}{\partial p_\phi} \,, & \dot{p}_\phi &= \mathcal{F}_\phi\,,
\end{align}
\end{subequations}
where ($r$,$\phi$,$p_r$,$p_\phi$) are the canonical variables in polar coordinates. (The construction of $H_{\rm eff}$ will be described in the next section.) Employing results from the \texttt{SEOBNRv5} model~\cite{Khalil:2023kep,Pompili:2023tna}, the radiation-reaction force ($\mathcal{F}_\phi$) is computed by summing over the PN GW modes (augmented with GSF information~\cite{vandeMeent:2023ols}) in a factorized form \cite{Damour:2008gu, Damour:2007xr, Pan:2010hz, Damour:2007yf,Pompili:2023tna}, which are used to obtain the inspiral-plunge modes after enhancing them by non-quasi-circular corrections~\cite{Damour:2002vi, Taracchini:2013rva, Bohe:2016gbl,Pompili:2023tna} during the plunge.

For the merger-ringdown part of the EOB waveform, we use instead a phenomenological ansatz \cite{Damour:2014yha,Bohe:2016gbl,Cotesta:2018fcv,Pompili:2023tna}, informed by NR and BH perturbation theory, as realized in the \texttt{SEOBNRv5} model~\cite{Pompili:2023tna}. The start of the merger-ringdown waveform is enforced to be at the peak of the $(2,2)$-mode amplitude. The gravitational polarizations can be written as $h_+ - i h_\times = \sum_{\ell,m} {}_{-2} Y_{\ell m} (\varphi, \iota) h_{\ell m} (t)$, where ${}_{-2} Y_{\ell m} (\varphi, \iota)$ are the -2 spin-weighted spherical 
harmonics, with $\varphi$ and $\iota$ being the azimuthal and polar angles to the observer, in the source frame. In the EOB approach, the inspiral-merger-ringdown $(\ell, m)$ modes are given by
\begin{align}\label{eq:modes}
  h_{\ell m} = 
  \begin{cases} 
    h_{\ell m}^{\rm insp-plunge}, & t<t_{\text {peak }}^{22}, \\
    h_{\ell m}^{\rm merg-RD}, & t>t_{\text {peak }}^{22}.
 \end{cases}
\end{align}
where $t_{\text {peak }}^{22}$ is the time at which the (2,2) mode has a peak, generally associated to the merger time. Such a time is suitably chosen to agree with the corresponding time in NR waveforms (see below).

\sec{PM-informed EOB Hamiltonian} We employ an effective Hamiltonian similar to that recently
introduced in the \texttt{SEOB-PM} scattering model~\cite{Buonanno:2024vkx}:
\begin{align}\label{eq:Heff}
  &H_{\rm eff}=\frac{Mp_\phi(g_{a_+}a_++g_{a_-}\delta a_-)}{r^3+a_+^2(r+2M)}\\
  &\qquad+\sqrt{A\left(\mu^2+\frac{p_\phi^2}{r^2}+(1+B_{\rm np}^{\rm Kerr})p_r^2
  +B_{\rm npa}^{\rm Kerr}\frac{p_\phi^2a_+^2}{r^2}\right)}\,,\nn
\end{align}
where $\delta=(m_1-m_2)/M$, while $B_{\rm np}^{\rm Kerr}=\chi_+^2u^2-2u$ and  $B_{\rm npa}^{\rm Kerr}=-{(1+2u)}/{[r^2+a_+^2(1+2u)]}$, 
where $u=M/r$ is the dimensionless PM counting parameter. In the probe limit $\nu\to0$, $H_{\rm eff}$ reduces to the Hamiltonian of a probe $\mu$
moving under the influence of a Kerr BH with mass $M$ and directed spin length $a_+$.
This Hamiltonian is determined by computing the scattering angle
and matching it to established PM results, 
but here we use only the conservative part of the angle
containing terms with even powers in the center-of-mass momentum $p_\infty=\mu\sqrt{\gamma^2-1}$,
where $\gamma=E_{\rm eff}/\mu>1$ for scattering trajectories.

Following Ref.~\cite{Buonanno:2024vkx}, the $\nu$-corrections with respect to the probe limit are built into the $A$-potential and
the gyro-gravitomagnetic factors as $A={(1-2u+\chi_+^2u^2+\Delta A)}/{[1+\chi_+^2u^2(2u+1)]}$
and $g_{a_\pm}={\Delta g_{a_\pm}/}{u^2}$. These respectively carry the even- and odd-in-spin corrections,
and are PM expanded up to a physical 5PM ($u^5$) order (see Table II in Ref.~\cite{Buonanno:2024vkx}):
\begin{align}\label{eq:pmExpand}
  \Delta A&=
  \sum_{n=2}^5u^n\Delta A^{(n)}+
  \Delta A^{\rm 4PN}\,, & 
  \Delta g_{a_\pm}&=
  \sum_{n=2}^5u^n\Delta g_{a_\pm}^{(n)}\,.
\end{align}
The $\gamma$-dependent coefficients $\Delta A^{(n)}$ and $\Delta g_{a_\pm}^{(n)}$
are series expanded in even powers of the spins, up to a highest quartic order at 5PM.
We lack an analytic 5PM term only in the non-spinning case, where the complete result
is not currently known (see Ref.~\cite{Driesse:2024xad} for the recently derived 1GSF conservative contribution).
Technically, as $\gamma=E_{\rm eff}/\mu \equiv H_{\rm eff}/\mu$, the Hamiltonian~\eqref{eq:Heff} is self-dependent. 
To produce an expression depending only on the canonical variables ($r$,$p_r$,$p_\phi$), 
we interpret $\gamma=H_{\rm Kerr}/\mu$ within these deformations,
plus whatever corrections are required in order to ensure the full Hamiltonian $H_{\rm EOB}$
is correct up to the desired PM order.
This procedure was used previously in the non-spinning case~\cite{Damour:2017zjx,Antonelli:2019ytb,Khalil:2022ylj},
and is fully described in the Supplemental Material.

An important subtlety within our Hamiltonian is the presence of non-local-in-time contributions (tails).
These imply a dependence on the full past history of the binary,
and thus distinguish between elliptic and hyperbolic (scattering) trajectories.
In the scattering Hamiltonian presented in Ref.~\cite{Buonanno:2024vkx},
tails are signaled by factors of $\log(\gamma^2-1)$,
which develops an imaginary part when $\gamma<1$ for bound orbits. 
To produce a real Hamiltonian, we therefore replace $\log(\gamma^2-1)\to\log(u)$ (see Supplemental 
Material for details).
We also include the 4PN non-spinning bound-orbit correction $\Delta A^{\rm 4PN}$ in Eq.~\eqref{eq:pmExpand},
\begin{align}\label{eq:PNcorrection}
  \Delta A^{\rm 4PN}&=u^4(\gamma^2-1)c_1+u^5(c_2+c_3\log u)\,,
\end{align}
ensuring the correct bound-orbit dynamics at 4PN order in the non-spinning case
(the numerical coefficients $c_i$ are provided in the Supplemental Material).
We verify our complete EOB Hamiltonian up to 4.5PN order~\cite{Damour:2014jta,Antonelli:2020aeb,Mandal:2022nty,Kim:2022pou,Levi:2014gsa}
by finding a suitable canonical transformation to its PN-expanded counterpart.\footnote{We thank Mohammed Khalil for providing us with suitable PN Hamiltonians to compare with.}
The non-spinning component is determined only up to quadratic order in eccentricity ($p_r^2$) in the tail integral,
as higher powers in eccentricity appear at lower PM orders.
Thus, we ensured that the 1PM--3PM (tail-free) non-spinning dynamics are unmodified by the presence of the 4PN correction~\eqref{eq:PNcorrection}.\footnote{As a spin-dependent bound-orbit correction would only arise at 5.5PN order, we choose not to include it.
}

\begin{figure}
  \vspace{-8pt}
  \includegraphics[width=.46\textwidth]{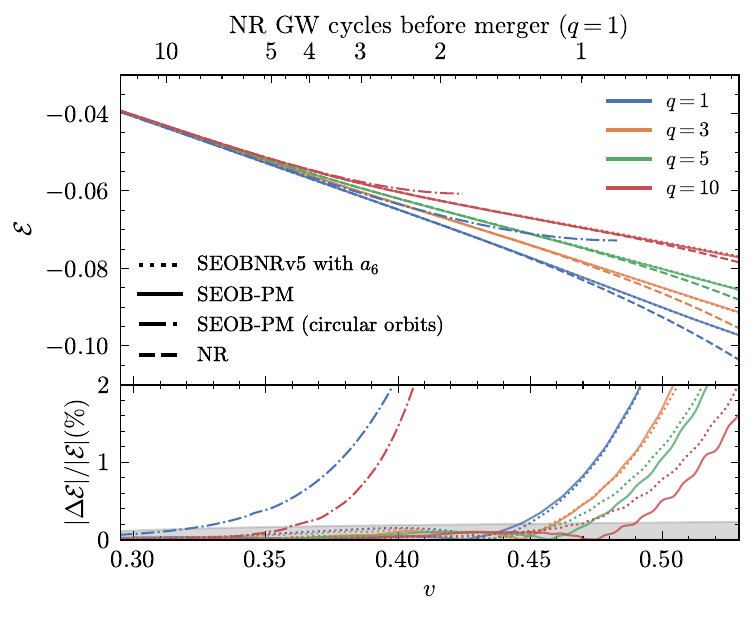}
  \captionsetup{skip=5pt}
  \caption{Non-spinning binding energy as a function of the (quasi-circular) 
velocity $v = (M\dot{\phi})^{2/3}$, for the (calibrated) \texttt{SEOBNRv5} with $a_6$ and (uncalibrated) \texttt{SEOB-PM} Hamiltonians (both 
along a circular orbit~\cite{Antonelli:2019ytb,Khalil:2022ylj} and inspiral) 
across different mass ratios $q=m_1/m_2$. The shaded region is an estimate of the NR uncertainty~\cite{Ossokine:2017dge}. The lower 
panel shows the fractional difference.}
  \label{fig:binding_nonspinning}
  \vspace{-8pt}
\end{figure}

\begin{figure*}
  \vspace{-8pt}
	\centering
	\subfloat{{\includegraphics[width=.46\linewidth]{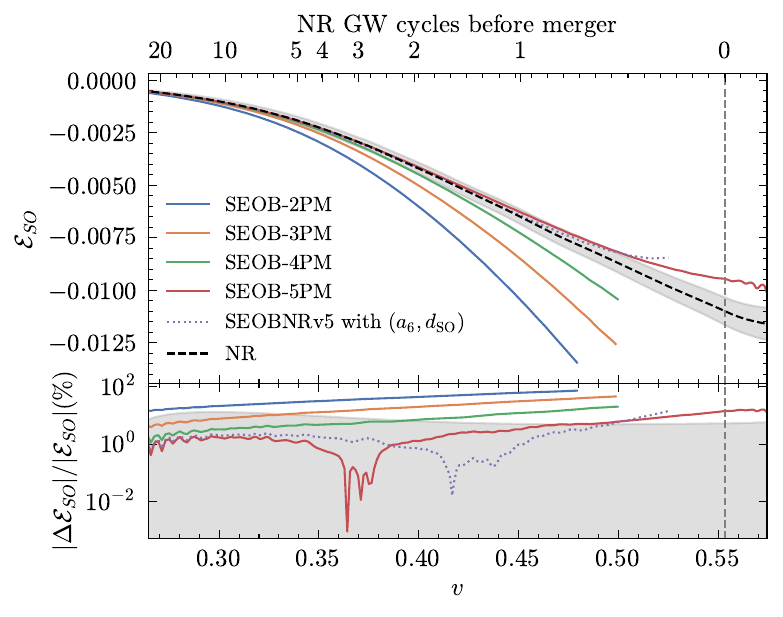} }} \hspace{0.5cm}
	\subfloat{{\includegraphics[width=.46\linewidth]{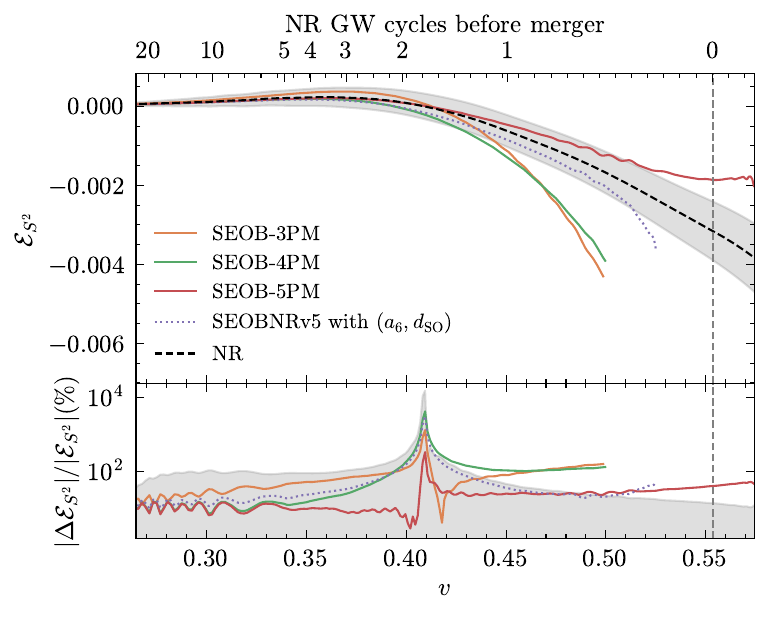} }}%
  \captionsetup{skip=5pt}
	\caption{Spin-orbit (left-hand panel) and spin-squared (right-hand panel) contributions to the binding energy, for an equal-mass BBH, as a function of the (quasi-circular) 
velocity $v = (M\dot{\phi})^{2/3}$, for the (calibrated) \texttt{SEOBNRv5} with ($a_6$,$d_{\rm SO}$) and (uncalibrated) \texttt{SEOB-PM} Hamiltonians at 
different PM orders. The vertical line represents the merger of the NR configuration (the one at the lowest velocity among those used),
with the number of GW cycles (top axis) referring to the same simulation. The shaded regions are estimates of the NR uncertainty~\cite{Ossokine:2017dge}.
The lower panel shows the absolute value of the fractional difference. The feature in the lower-right panel around $v \sim 0.4$ is due to a zero-crossing.}
	\label{fig:binding_spinning}
  \vspace{-8pt}
\end{figure*}

Finally, let us comment on the appearance of special functions in $H_{\rm eff}$.
Starting at 3PM order we encounter the combination ${{\rm arccosh}(\gamma)}/{\sqrt{\gamma^2-1}}$.
As ${\rm arccosh}(\gamma)$ and $\sqrt{1-\gamma^2}$ are both imaginary valued when $\gamma<1$,
we find it convenient to replace this combination by ${\arccos(\gamma)}/{\sqrt{1-\gamma^2}}$,
which has the same small-velocity expansion for scattering kinematics.
At 4PM order we then encounter logarithms, dilogarithms (${\rm Li}_2$) and elliptic functions (K/E)
of the first and second kind, all evaluated as functions of $\gamma$.
In this case, we also find it convenient to introduce the inverse tangent integral
${\rm Ti}_2(x):=\int_0^x({\rm d}t/t)\arctan t$, analogously to what is done above. 
Fast numerical routines exist for evaluating all of these functions in Cython~\cite{Behnel:2011ddo},
and this leads to an efficient numerical evaluation within \texttt{pySEOBNR}~\cite{Mihaylov:2023bkc}.

\sec{Comparing \texttt{SEOB-PM} and NR binding energies during the inspiral} In EOB models one has access to the 
binary's dynamics, which enables testing their accuracy by comparing (gauge-invariant) dynamical quantities
such as the binding energy~\cite{Damour:2011fu, LeTiec:2011dp, Nagar:2015xqa, Ossokine:2017dge, Khalil:2022ylj} 
and periastron advance~\cite{LeTiec:2011bk,Hinderer:2013uwa}. As \texttt{SEOBNR-PM}'s essential new feature
is its PM-informed \texttt{SEOB-PM} Hamiltonian, 
the binding energy is a particularly relevant quantity to compare with NR data. 
Previous comparisons in the non-spinning case~\cite{Antonelli:2019ytb,Khalil:2022ylj}
have focused on the binding energy computed for circular orbits (i.e., ignoring radiation-reaction effects), 
although Ref.~\cite{Antonelli:2019ytb} investigated the effect of neglecting dissipation (see Fig.~6 therein). 
We instead compute the (dimensionless) binding energy by evaluating ${\cal E}=(H_{\mathrm{EOB}}-M)/\mu$ along the inspiraling dynamics, and  
compare with NR--binding-energy data from Ref.~\citep{Ossokine:2017dge}. Fig.~\ref{fig:binding_nonspinning} shows 
the EOB and NR non-spinning binding energies as a function of the (quasi-circular) velocity parameter 
$v=(M \dot \phi )^{1/3}$, for \texttt{SEOBNRv5} with $a_6$ and  \texttt{SEOB-PM} for circular orbits and along an inspiral. We stress that for the former, a 5PN-unknown parameter 
($a_6$) in the $A$-potential has been calibrated against 18 non-spinning simulations (see below). Both models show excellent agreement with 
NR during most of the inspiral, with errors within the NR uncertainty (represented by the gray region) until around 1 GW cycle before merger.
The (uncalibrated) \texttt{SEOB-PM} maintains agreement within NR error up to slightly higher velocities for higher mass ratios, and  
it has much better agreement than when computed on circular orbits~\cite{Antonelli:2019ytb,Khalil:2022ylj}.

We also extract different spin contributions to the binding energy
by combining results from NR simulations for various equal-mass spin combinations~\cite{Dietrich:2016lyp, Ossokine:2017dge}:
${\cal{E}}_{\mathrm{SO}}=-\frac{1}{6}{\cal{E}}(-0.6,0)+\frac{8}{3}{\cal{E}}(0.3,0)-2{\cal{E}}(0,0)-\frac{1}{2}{\cal{E}}(0.6,0)+\mathcal{O}(S^3)$ and
${\cal{E}}_{\mathrm{S^2}}=\frac{3}{2}{\cal{E}}(-0.6,0)-2{\cal{E}}(0,0)+\frac{3}{2}{\cal{E}}(0.6,0)-{\cal{E}}(0.6,-0.6)+\mathcal{O}(S^3)$,
where ${\cal{E}}(\chi_1, \chi_2)$ denotes the binding energy in a simulation with dimensionless spins $\chi_i$.
In Fig.~\ref{fig:binding_spinning} we illustrate the spin-orbit and spin-squared contributions for an equal-mass BBH to the binding energy for the (uncalibrated) \texttt{SEOB-PM}  at different PM orders, as compared with NR and with the (calibrated) \texttt{SEOBNRv5} with ($a_6$, $d_{\rm SO}$). Despite not being calibrated to NR, \texttt{SEOB-PM} shows excellent agreement with the NR results, with a clear convergence toward the NR prediction, as more PM corrections are included. Its accuracy is somewhat better than \texttt{SEOBNRv5}, despite the latter model using a Hamiltonian calibrated in the non-spinning ($a_6$) and spin-orbit coupling ($d_{\rm SO}$) sector (see below).

\sec{Calibration to numerical-relativity waveforms} As discussed, the accuracy of EOB inspiral-merger-ringdown waveforms can be enhanced through calibration to NR simulations. For the inspiral-plunge stage, this is generally achieved by introducing in the Hamiltonian high-order (still unknown) PN terms, whose coefficients are tuned to NR, and fitting the time of merger (i.e., the (2,2)-mode's peak time) to NR. In the \texttt{SEOBNRv5} model~\cite{Pompili:2023tna}, which was built integrating PN results in the Hamiltonian, three calibration parameters were employed: $(\Delta t_{\rm NR}, a_6,d_{\rm SO})$. The parameter $\Delta t_{\rm NR}$ is defined by $t_{\text {peak }}^{22}=t_{\mathrm{ISCO}}+\Delta t_{\mathrm{NR}}$ (see also Eq.~(\ref{eq:modes})), where $t_{\rm{ISCO}}$ is the time at which $r = r_{\rm{ISCO}}$, with $r_{\rm{ISCO}}$ the radius of the Kerr innermost stable circular orbit (ISCO)~\cite{Bardeen:1972fi} with the mass and spin of the remnant BH, as given by NR fitting formula ~\cite{Jimenez-Forteza:2016oae, Hofmann:2016yih}. The parameter $a_6$ is a 5PN correction to the $A$-potential and $d_{\rm SO}$ is a 4.5PN correction in the gyro-gravitomagnetic coefficients.\footnote{More specifically, \texttt{SEOBNRv5}'s Hamiltonian employs all known 5PN coefficients in the $A$- and $D$-potentials (except for the partially known 5PN local part of the $A$-potential that is fully replaced by the $a_6$ calibration coefficient), and the 5.5PN terms in the $Q$-potential. Although the 4.5PN spin-orbit couplings are known, Ref.~\cite{Pompili:2023tna} did not use them (except for the 3.5PN) because the authors showed that having a spin-orbit calibration parameter at 5.5PN  (instead of 4.5PN) is not very effective.}
Here, for the \texttt{SEOBNR-PM} model, we do not calibrate high-order PN terms in the non-spinning and spin sectors of the Hamiltonian (\ref{eq:Heff}), but we calibrate only the merger's time through $\Delta t_{\rm NR}$.
In future work, we will explore 
NR calibrations tailored to the particular structure of the PM terms. Henceforth, we compare the PM-informed model with several versions of the most recent PN-GSF--informed \texttt{SEOBNRv5}, with and without calibration.

Waveform accuracy is often quantified in terms of the mismatch $\mathcal{M}$, defined as $1$ minus the overlap between the normalized waveforms, maximized over a relative time and phase shift:
\begin{equation}
	\label{eq:mismatch}
	\mathcal{M}= 1 - \max _{\phi_0, t_0} \frac{\left(h_1 \mid h_2\right)}{\sqrt{\left(h_1 \mid h_1\right)\left(h_2 \mid h_2\right)}}.
\end{equation}
The overlap is a noise-weighted inner product \cite{Sathyaprakash:1991mt, Finn:1992xs} 
$\left(h_1 \mid h_2\right) \equiv 4 \operatorname{Re} \int_{f_l}^{f_h} df \tilde{h}_1(f) \tilde{h}_2^*(f) / S_n(f) $, where $\tilde h(f)$ is the Fourier transform of the time-domain signal, the ${}^*$ superscript indicates complex conjugation, and $S_n(f)$ is the power spectral density of the detector noise, which we assume to be the design zero-detuned high-power noise curve of Advanced LIGO~\cite{Barsotti:2018}.

To calibrate the \texttt{SEOBNR-PM} model, we closely follow the
procedure outlined in Refs.~\cite{Bohe:2016gbl, Pompili:2023tna,Mihaylov:2023bkc}. This
procedure essentially consists of determining values of the
calibration parameters that minimize a combination of the mismatch and
the difference in merger time (defined as the peak of the $(2,2)$-mode
amplitude) between EOB and NR waveforms with the same physical
parameters ($q, \chi_1, \chi_2$). This is carried out in a Bayesian
fashion using the \texttt{Bilby}~\cite{Ashton:2018jfp} package,
  and the \texttt{pySEOBNR} code~\cite{Mihaylov:2023bkc}. Finally, we
  interpolate the best-fit values for each NR simulation across the
  ($q, \chi_1, \chi_2$) parameter space. As said, in our
  \texttt{SEOBNR-PM} model, we only calibrate the $\Delta t_{\rm NR}$
  parameter (see the Supplemental Material for its expression) using 441 NR simulations
  of aligned-spin BBHs produced with the pseudo-Spectral Einstein code
  of the Simulating eXtreme Spacetimes (SXS)
  Collaboration~\cite{Boyle:2019kee, Mroue:2013xna, Hemberger:2013hsa,
    Scheel:2014ina, Lovelace:2014twa, Chu:2015kft, Blackman:2015pia,
    Abbott:2016nmj, Abbott:2016apu, Bohe:2016gbl, Lovelace:2016uwp,
    Blackman:2017dfb, Varma:2018mmi, Varma:2019csw, Yoo:2022erv},
  which were also employed in Ref.~\cite{Pompili:2023tna} for the
  \texttt{SEOBNRv5} model. They cover mass ratios $q=m_1/m_2$ from
$1$ to $20$ in the non-spinning limit, and dimensionless spin values
going from $-0.998 \leq \chi_i \leq 0.998$ for $q=1$ to $-0.5 \leq
\chi_1 \leq 0.5,~\chi_2=0$ for $q=15$.

\begin{figure}[t]
  \vspace{-6pt}
  \includegraphics[width=.46\textwidth]{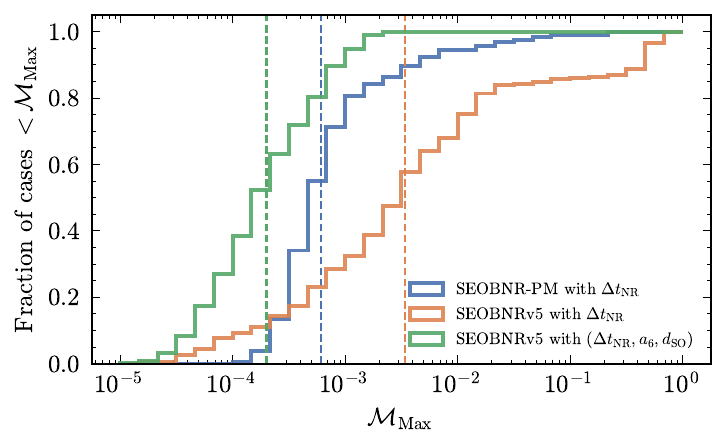}
  \captionsetup{skip=5pt}
  \caption{Cumulative maximum mismatch over the binary's total-mass range $10 M_{\odot} \leq M \leq 300 M_{\odot}$ for the (calibrated) \texttt{SEOBNR-PM} and 
\texttt{SEOBNRv5} models. The study uses 441 SXS NR waveforms, and focuses on the $(\ell,m)=(2,2)$ mode. The vertical dashed lines indicate the medians of the 
mismatch distributions.
  }
  \label{fig:mismatch}
  \vspace{-6pt}
\end{figure}

\sec{\texttt{SEOBNR-PM} waveform-model performance} To assess the
accuracy of the waveform model, we compute its mismatch against the
set of 441 SXS NR simulations, and compare its performance to the
\texttt{SEOBNRv5} $(\Delta t_{\rm NR}, a_6,d_{\rm SO})$ model, as well
as, to a version of \texttt{SEOBNRv5} calibrated only via $\Delta
t_{\mathrm{NR}}$. Fig.~\ref{fig:mismatch} illustrates the cumulative maximum 
mismatch against the NR simulations over
the binary's total-mass range $10 M_{\odot} \leq M \leq 300 M_{\odot}$, for the
$(\ell,m)=(2,2)$ mode. The overall mismatch of \texttt{SEOBNR-PM}
against NR falls roughly between that of the two \texttt{SEOBNRv5}
variations, with a median value $\mathcal{M}_{\mathrm{median}}\sim 6.1
\times 10^{-4}$. This represents a remarkably good
agreement. When tuning only $\Delta t_{\rm NR}$, we observe that the accuracy of
both \texttt{SEOBNR-PM} and \texttt{SEOBNRv5} tends to degrade for
configurations with large positive spins. 
This results in a tail of high-mismatch cases above $\sim 1\%$, more pronounced for \texttt{SEOBNRv5}, which includes spin-orbit (3.5PN), spin-square (4PN), and spin-cube (3.5PN) effects at a lower PN order than \texttt{SEOBNR-PM}, which includes spin terms up to 5PM order~\footnote{When using the \texttt{SEOBNRv5} model with 4.5PN spin-orbit couplings and calibrate $\Delta t_{\rm NR}$, we find a better 
agreement than the model with 3.5PN, but \texttt{SEOBNR-PM} performs still better.}. Resumming the PM-EOB potentials and introducing calibration parameters could greatly improve \texttt{SEOBNR-PM}'s accuracy for these cases, similar to the calibrated \texttt{SEOBNRv5}. 
We leave this important work to the future.

\sec{Conclusions}
In this Letter, we took advantage of the flexible and efficient \texttt{pySEOBNR} code~\cite{Mihaylov:2023bkc} and 
recent prediction for the scattering angle in the EOB formalism~\cite{Buonanno:2024vkx} to build the first inspiral-merger-ringdown 
EOB waveform model (\texttt{SEOBNR-PM}) for aligned-spin BHs that uses a PM-informed Hamiltonian (i.e., expanded in $G$, but 
at all orders in the velocity). Importantly, we found that the \texttt{SEOB-PM} non-spinning binding energy, computed along an inspiraling trajectory, at 4PM, and its spin-orbit and spin-spin contributions through 5PM, agree remarkably well with the NR data up to 1 GW cycle before merger (see Figs.~\ref{fig:binding_nonspinning} and ~\ref{fig:binding_spinning}). The agreement is comparable and in some cases better than \texttt{SEOBNRv5}, which however was calibrated to NR results~\cite{Pompili:2023tna}. 
Furthermore, we calibrated \texttt{SEOBNR-PM} to 441 NR simulations provided by the SXS Collaboration~\cite{Boyle:2019kee, Mroue:2013xna, Hemberger:2013hsa, Scheel:2014ina, Lovelace:2014twa, Chu:2015kft, Blackman:2015pia, Abbott:2016nmj, Abbott:2016apu, Bohe:2016gbl, Lovelace:2016uwp, Blackman:2017dfb, Varma:2018mmi, Varma:2019csw,  Yoo:2022erv} by tuning the (2,2)-mode's peak time (i.e., $\Delta t_{\rm NR}$), and found a median mismatch 
lower than \texttt{SEOBNRv5}, when the latter is similarly calibrated to NR (see Fig.~\ref{fig:mismatch}).
For now, without optimization, the \texttt{SEOBNR-PM}'s evaluation time is 
an order of magnitude slower than \texttt{SEOBNRv5}.

Considering the recent attention to the two-body gravitational-scattering
problem in quantum-field theory,  
with a slew of new results produced~\cite{Bern:2021dqo,Bern:2021yeh,Damgaard:2023ttc,
Dlapa:2021vgp,Dlapa:2023hsl,Jakobsen:2023ndj,Jakobsen:2023pvx}, 
we see the development of the \texttt{SEOBNR-PM} model as a watershed moment --- the first true application of these methods to
an astrophysically relevant inspiral-merger-ringdown waveform model. 
Yet, this is only a first step. Given the relevant progress at 5PM~\cite{Driesse:2024xad}, we hope 
to incorporate the complete 5PM scattering angle into our effective
Hamiltonian in the near future. Recent results separating the local from non-local parts of the 4PM angle~\cite{Dlapa:2024cje}
will likely be crucial for achieving good agreement with NR for highly elliptic bound systems ---
ultimately, this may be the \texttt{SEOBNR-PM} model's \emph{raison d'\^{e}tre}.  
In light of the progress in PM fluxes~\cite{Herrmann:2021lqe,Herrmann:2021tct,DiVecchia:2021bdo,Heissenberg:2021tzo,Bjerrum-Bohr:2021din,Damgaard:2021ipf,Mougiakakos:2021ckm,Manohar:2022dea,Dlapa:2022lmu,Jakobsen:2021smu,Jakobsen:2021lvp,Jakobsen:2023hig}, PM corrections could also be fed into the EOB 
radiation-reaction forces and gravitational modes. The \texttt{SEOB-PM} Hamiltonian and fluxes will also need to be extended 
to the astrophysically relevant precessing-spin case. We leave these tantalizing prospects for future work.

\sec{Acknowledgments}
We are grateful to Zvi Bern, Gustav Uhre Jakobsen, Mohammed Khalil, Jan Plefka and
Jan Steinhoff for valuable discussions and comments on this Letter,
and to Raffi Enficiaud for his assistance with scientific computing.
The work of G.M. and R.P. was supported by the Deutsche Forschungsgemeinschaft
(DFG) Projektnummer 417533893/GRK2575 ``Rethinking Quantum Field Theory''.

\section*{Supplemental Material}

\appendix

\section{PM-informed effective Hamiltonian in the EOB approach}

Here we present the effective Hamiltonian $H_{\rm eff}$~\eqref{eq:Heff},
which depends on the $A$-potential and gyro-gravitomagnetic factors:
\begin{align}
A&=\frac{1-2u+\chi_+^2u^2+\Delta A}{1+\chi_+^2u^2(2u+1)}\,, &
g_{a_\pm}&=\frac{\Delta g_{a_\pm}}{u^2}\,.
\end{align}
These are PM-expanded in Eq.~\eqref{eq:pmExpand}.
Corrections to the $A$-potential incorporate even-in-spin PM corrections:
\begin{align}\label{eq:deltaA}
  \Delta A^{(n)}=
  \sum_{s=0}^{\lfloor (n-1)/2 \rfloor}\sum_{i=0}^{2s}
  \alpha^{(n)}_{(2s-i,i)}\delta^{\sigma(i)}\chi_+^{2s-i}\chi_-^{i}\,,
\end{align}
where $\sigma(i)=0,1$ if $i$ is even or odd, respectively.
The gyro-gravitomagnetic factors incorporate odd-in-spin PM corrections:
\allowdisplaybreaks
\begin{subequations}
  \begin{align}
    \Delta g_{a_+}^{(n)}&=\!\!
    \sum_{s=0}^{\lfloor (n-2)/2 \rfloor}\sum_{i=0}^s
    \alpha^{(n)}_{(2(s-i)+1,2i)}\chi_+^{2(s-i)}\chi_-^{2i}\,,\\
    \Delta g_{a_-}^{(n)}&=\!\!
    \sum_{s=0}^{\lfloor (n-2)/2 \rfloor}\sum_{i=0}^s
    \alpha^{(n)}_{(2(s-i),2i+1)}\chi_+^{2(s-i)}\chi_-^{2i}\,.
  \end{align}
\end{subequations}
The dimensionless parameters $\alpha^{(n)}_{(i,j)}$
are functions of $\gamma=E_{\rm eff}/\mu$ and the symmetric mass ratio $\nu$.

We now provide the deformation coefficients required in order to fully specify the effective Hamiltonian
for non-spinning configurations.
For the full spinning, we refer the interested reader to the ancillary file attached to the \texttt{arXiv}
submission of this Letter that contains the complete  EOB and effective Hamiltonians.
Firstly, the 2PM non-spinning deformation is
\begin{align}
\alpha^{(2)}_{(0,0)}=\frac{3(\Gamma-1)(5\gamma^2-1)}{2\gamma^2\Gamma}\,,
\end{align}
where $\Gamma=\sqrt{1+2\nu(\gamma-1)}$ is the dimensionless total energy.
Next, at 3PM order we require
\begin{widetext}
\begin{align}
\alpha^{(3)}_{(0,0)}=
\frac{9(\Gamma\!-\!1)\left(30 \gamma ^4\!-\!31 \gamma ^2\!+\!5\right)+2\nu\left(-214 \gamma ^5\!+\!270 \gamma ^4\!+\!323 \gamma ^3\!-\!279
   \gamma ^2\!-\!145 \gamma \!+\!45\right)}{6 \gamma ^2 \left(\gamma ^2-1\right) \Gamma ^2}
+\frac{4\nu\left(4 \gamma ^4\!-\!12 \gamma ^2\!-\!3\right)}{\gamma ^2\Gamma ^2}
\frac{\arccos\gamma}{\sqrt{1-\gamma ^2}}\,.
\end{align}
Here, we have replaced ${\rm arccosh}(\gamma)/\sqrt{\gamma^2-1}$ by $\arccos(\gamma)/\sqrt{1-\gamma^2}$.
Finally, at 4PM order we encounter
\allowdisplaybreaks
\begin{align}
&\alpha^{(4)}_{(0,0)}=
\frac{7\nu\left(380 \gamma ^2+169\right)}{8 (\gamma -1) \gamma ^2 \Gamma ^3}{\rm E}^2\left(\frac{\gamma -1}{\gamma+1}\right)
+\frac{\left(1200 \gamma ^2+2095 \gamma +834\right) \nu}{4 \gamma ^2 \left(\gamma ^2-1\right) \Gamma ^3}{\rm K}^2\left(\frac{\gamma -1}{\gamma +1}\right)\\
&+\frac{\left(-1200 \gamma ^3-2660 \gamma ^2-2929 \gamma -1183\right) \nu}{4 \gamma ^2 \left(\gamma ^2-1\right) \Gamma
   ^3}{\rm E}\left(\frac{\gamma -1}{\gamma +1}\right){\rm K}\left(\frac{\gamma -1}{\gamma +1}\right)\nn\\
&+\frac{\left(-25 \gamma ^6+30 \gamma ^4+111 \gamma ^2+20\right) \nu}{\gamma ^2 \Gamma ^3} \text{Li}_2\left(\frac{1-\gamma }{1+\gamma}\right)+\frac{(\gamma +1) \left(25 \gamma ^5-25 \gamma ^4-5 \gamma ^3+65 \gamma
   ^2+64 \gamma +12\right) \nu}{2 \gamma ^2 \Gamma ^3}\text{Li}_2\left(\frac{\gamma -1}{\gamma +1}\right)\nn\\
&+\frac{\left(35 \gamma ^4+120
   \gamma ^3+90 \gamma ^2+152 \gamma +27\right) \nu}{2 \gamma ^2 \Gamma ^3}\log ^2\left(\frac{\gamma +1}{2}\right)
-\frac{4\left(2\gamma ^2-3\right)\left(15\gamma ^2-15 \gamma +4\right)\nu}{\gamma(\gamma +1)\Gamma ^3}
\frac{{\rm Ti}_2\left(\sqrt{\frac{1-\gamma}{1+\gamma}}\right)}{\sqrt{1-\gamma^2}}\nn\\
&+\frac{\left(2 \gamma ^2-3\right)^2 \left(35 \gamma ^4-30 \gamma ^2+11\right) \nu}{8 \left(\gamma ^2-1\right)^3 \Gamma ^3} \arccos^2\gamma
+\frac{2 \left(75 \gamma ^6-140
   \gamma ^4-283 \gamma ^2-852\right) \nu}{3 \gamma  \left(\gamma ^2-1\right) \Gamma ^3}  \log (\gamma )\nn\\
&+\frac{\left(210 \gamma ^6-552 \gamma ^5+339 \gamma ^4-912 \gamma ^3+3148 \gamma ^2-3336 \gamma +1151\right) \nu  }{12 \gamma ^2 \left(\gamma ^2-1\right) \Gamma ^3}\log \left(\frac{u}{4}\right)\nn\\
&+\bigg(\frac{\left(-35 \gamma ^4-60 \gamma ^3+150 \gamma ^2-76 \gamma +5\right)\nu }{2 \gamma ^2
   \Gamma ^3}\log \left(\frac{u}{4}\right)\nn\\
&\qquad+\frac{\left(-75 \gamma ^7+416 \gamma ^5+612 \gamma ^4+739 \gamma ^3+136 \gamma ^2+2520 \gamma +152\right) \nu }{3 \gamma ^2 \left(\gamma ^2-1\right) \Gamma ^3}\bigg) \log \left(\frac{\gamma
   +1}{2}\right)\nn\\
&+\bigg(
  \frac{\left(-420 \gamma ^9+96 \gamma ^8-48 \gamma ^7+5328 \gamma ^6-5279 \gamma
   ^5-1584 \gamma ^4+7142 \gamma ^3-9360 \gamma ^2+3453 \gamma +720\right) \nu}{12 \gamma ^2 (\gamma^2-1)^{2} \Gamma ^3}\nn\\
&\qquad-\frac{48\left(7 \gamma ^2-5\right) \left(4 \gamma ^4-12 \gamma ^2-3\right) (\Gamma -1) \nu }{12 \gamma ^2 (\gamma^2-1) \Gamma ^3}\nn\\
&\qquad-\frac{\left(2 \gamma ^2-3\right) \left(35 \gamma ^4-30 \gamma ^2+11\right) \nu}{4 \gamma(1-\gamma ^2)\Gamma ^3}\log \left(\frac{u}{4}\right)+\frac{4 \left(2 \gamma ^2-3\right)
   \left(15 \gamma ^2+2\right) \nu  }{\gamma(1-\gamma ^2)\Gamma ^3}\log \left(\frac{\gamma +1}{2}\right)\bigg)\frac{\arccos\gamma}{\sqrt{1-\gamma^2}}\nn\\
&+(\Gamma -1) \bigg(\frac{5115 \gamma ^8-9537 \gamma ^6+5657 \gamma ^4-1115 \gamma ^2+72}{16 \gamma ^4 \left(\gamma ^2-1\right)^2 \Gamma ^3}\nn\\
&\qquad+\frac{\left(8159 \gamma ^8-3136 \gamma ^7-23601 \gamma ^6-3360 \gamma ^5+15409 \gamma ^4+4000
\gamma ^3-1995 \gamma ^2+108\right) \nu }{24 (\gamma -1) \gamma ^4 (\gamma +1)^2 \Gamma ^3}\bigg)\nn\\
&+\frac{\nu}{144 \gamma ^9 \left(\gamma ^2-1\right)^2 \Gamma ^3}\bigg(-600 \pi ^2 \gamma ^{17}+3600 \gamma ^{16}+480 \left(9+4 \pi ^2\right) \gamma ^{15}+2 \left(720 \pi ^2-28843\right)
\gamma ^{14}+\left(36759-5136 \pi ^2\right) \gamma ^{13}\nn\\
&\qquad+\left(44698-1056 \pi ^2\right) \gamma ^{12}+\left(6624 \pi ^2-43235\right) \gamma ^{11}+\left(7702-2208 \pi ^2\right) \gamma ^{10}-5 \left(2155+504 \pi ^2\right) \gamma^9\nn\\
&\qquad+2 \left(23947+912 \pi ^2\right) \gamma ^8-\left(45605+288 \pi ^2\right) \gamma ^7+12701 \gamma ^6+648 \gamma ^5-1471 \gamma ^4+207 \gamma ^2-45 \bigg)\,,\nn
\end{align}
\end{widetext}
where ${\rm Ti}_2(x):=\int_0^x({\rm d}t/t)\arctan t$ is the inverse tangent integral. 
This expression also includes $\log(u)$,
which is the logarithmic bound-orbit counterpart to $\log(\gamma^2-1)$.
The $\log(u)$ can be identified 
by taking the bound circular-orbit limit (at leading-PN order) of $\log(1-\gamma^2)$,
i.e.~the real part of $\log(\gamma^2-1)$,
ensuring that the coefficient is real for $\gamma<1$.
Finally, to complete the non-spinning dynamics we require the 4PN correction coefficient $\Delta A^{\rm 4PN}$~\eqref{eq:PNcorrection}.
This includes three numerical coefficients:
\begin{subequations}
\begin{align}
  &c_1=\frac{\nu}{15}(-1411\!+\!296 \gamma_{\rm E} \!-\!1328 \log2\!+\!2187 \log3)\,,\\
  &c_2=\frac{9 \nu ^3}{4}\!+\!\frac{\nu^2}{192} \left(615 \pi ^2-16408\right)\\
  &\!+\!\nu  \left(-\frac{51187}{360}\!+\!\frac{136 \gamma_{\rm E} }{3}\!+\!\frac{1571 \pi ^2}{6144}
  -\frac{856 \log2}{15}\!+\!\frac{729 \log3}{5}\right)\!,\nn\\
  &c_3=-\frac{68\nu}{3}\,,
\end{align}
\end{subequations}
with $\gamma_{\rm E}\approx0.557$ Euler's constant.

\section{Interpreting $\gamma$ within $\alpha^{(n)}_{(i,j)}$}

The effective Hamiltonian $H_{\rm eff}$~\eqref{eq:Heff} contains deformation parameters $\alpha^{(n)}_{(i,j)}$ that depend
on $\gamma=E_{\rm eff}/\mu$ (i.e., the Hamiltonian technically depends on itself).
In principle, one should therefore solve for $E_{\rm eff}=H_{\rm eff}$ as a function of the kinematic variables
($r$,$p_r$,$p_\phi$);
however, given the highly non-trivial dependence on $\gamma$ on the right-hand side of Eq.~\eqref{eq:Heff},
not to mention the appearance of special functions including dilogarithms and elliptics,
such an approach is not practical.
While one might also try solving for the Hamiltonian numerically at a given phase-space point,
this approach will not lead to an efficient implementation within \texttt{pySEOBNR}.

Instead, building on the approach taken in Ref.~\cite{Antonelli:2019ytb,Khalil:2022ylj} for the non-spinning case,
we interpret $\gamma$ as the effective energy expanded only up to whatever PM and/or spin order is necessary
to ensure that the complete resummed Hamiltonian $H_{\rm EOB}= M\, \sqrt{1 + 2 \nu (H_{\rm eff}/\mu -1) }$
is consistent with the known PM and spin results.
The effective Hamiltonian may be perturbatively expanded as
\begin{align}
  \label{eq:gamma}
  \gamma=\gamma_{\rm Kerr}+\sum_{n\geq2}\sum_{s\geq0}\Delta_{(s)}^{(n)}(\gamma_{\rm Kerr})
\end{align}
with deformations $\Delta_{(s)}^{(n)}(\gamma_{\rm Kerr})$, characterized by PM order $n$ and spin order $s$,
depending on $\gamma_{\rm Kerr}=H_{\rm Kerr}/\mu$.
The Kerr Hamiltonian here is
\begin{align}\label{eq:HKerr}
  &H_{\rm Kerr}=\frac{2Mp_\phi a}{r^3+a^2(r+2M)}\\
  &\quad+\sqrt{A^{\rm Kerr}\left(\mu^2+\frac{p_\phi^2}{r^2}+(1+B_{\rm np}^{\rm Kerr})p_r^2
  +B_{\rm npa}^{\rm Kerr}\frac{p_\phi^2a^2}{r^2}\right)}\,,\nn
\end{align}
where we identify the total spin $a=a_+$ in the EOB.
The three functions appearing in the Kerr Hamiltonian are 
\begin{subequations}
\begin{align}
  A^{\rm Kerr}&=\frac{1-2u+\chi^2u^2}{1+\chi^2u^2(2u+1)}\,,\\
  B_{\rm np}^{\rm Kerr}&=\chi^2u^2-2u\,,\\
  B_{\rm npa}^{\rm Kerr}&=-\frac{1+2u}{r^2+a^2(1+2u)}\,.
\end{align}
\end{subequations}
For our purposes, expressions for the following five deformations (up to 3PM order) are sufficient:
\begin{subequations}\label{eq:HeffExpand}
\begin{align}
  \Delta_{(0)}^{(2)}&=
  \frac{u^2}{2}\alpha^{(2)}_{(0,0)}(\gamma_{\rm Kerr})\gamma_{\rm Kerr}\,,\\
  \Delta_{(1)}^{(2)}&=
  \ell u^3 \left((\alpha^{(2)}_{(1,0)}(\gamma_{\rm Kerr})-2)a_++\alpha^{(2)}_{(0,1)}(\gamma_{\rm Kerr})\delta a_-\right),\\
  \Delta_{(0)}^{(3)}&=
  \frac{u^3}{2}\left(2\alpha^{(2)}_{(0,0)}(\gamma_{\rm Kerr})+\alpha^{(3)}_{(0,0)}(\gamma_{\rm Kerr})\right)\gamma_{\rm Kerr}\,,\\
  \Delta_{(1)}^{(3)}&=
  \ell u^4 \left(\alpha^{(3)}_{(1,0)}(\gamma_{\rm Kerr})a_++\alpha^{(3)}_{(0,1)}(\gamma_{\rm Kerr})\delta a_-\right)\!,\\
  \Delta_{(2)}^{(3)}&=
  \frac{u^3}{2}\left(\alpha^{(3)}_{(2,0)}(\gamma_{\rm Kerr})a_+^2+\alpha^{(3)}_{(0,2)}(\gamma_{\rm Kerr})a_-^2\right.\nn\\
    &\qquad\qquad\qquad+\left.\alpha^{(3)}_{(1,1)}(\gamma_{\rm Kerr})\delta a_+a_-\right)\gamma_{\rm Kerr}\,,
\end{align}
\end{subequations}
Within a given coefficient $\alpha_{(i,j)}^{(n)}$ we use whatever deformations in Eq.~\eqref{eq:HeffExpand} are required.
The explicit coefficient-by-coefficient replacements used in our model are
\begin{align}
    \alpha^{(2)}_{(0,0)}(\gamma)&: & \gamma&\to\gamma_{\rm Kerr}
    +\Delta^{(2)}_{(0)}+\Delta^{(2)}_{(1)}+\Delta^{(3)}_{(1)}+\Delta^{(3)}_{(2)}\,,\nn\\
    \alpha^{(3)}_{(0,0)}(\gamma)&: & \gamma&\to\gamma_{\rm Kerr}+
    \Delta^{(2)}_{(1)}\,,\nn\\
    \alpha^{(2)}_{(1,0)}(\gamma)&: & \gamma&\to\gamma_{\rm Kerr}
    +\Delta^{(2)}_{(0)}\!+\!\Delta^{(2)}_{(1)}\!+\!\Delta^{(3)}_{(0)}\!+\!\Delta^{(3)}_{(1)}\!+\!\Delta^{(3)}_{(2)}\,,\nn\\
    \alpha^{(2)}_{(0,1)}(\gamma)&: & \gamma&\to\gamma_{\rm Kerr}
    +\Delta^{(2)}_{(0)}\!+\!\Delta^{(2)}_{(1)}\!+\!\Delta^{(3)}_{(0)}\!+\!\Delta^{(3)}_{(1)}\!+\!\Delta^{(3)}_{(2)}\,,\nn\\
    \alpha^{(3)}_{(1,0)}(\gamma)&: & \gamma&\to\gamma_{\rm Kerr}
    +\Delta^{(2)}_{(0)}+\Delta^{(2)}_{(1)}\,,\nn\\
    \alpha^{(3)}_{(0,1)}(\gamma)&: & \gamma&\to\gamma_{\rm Kerr}
    +\Delta^{(2)}_{(0)}+\Delta^{(2)}_{(1)}\,,\nn\\
    \alpha^{(3)}_{(2,0)}(\gamma)&: & \gamma&\to\gamma_{\rm Kerr}
    +\Delta^{(2)}_{(0)}+\Delta^{(2)}_{(1)}\,,\nn\\
    \alpha^{(3)}_{(0,2)}(\gamma)&: & \gamma&\to\gamma_{\rm Kerr}
    +\Delta^{(2)}_{(0)}+\Delta^{(2)}_{(1)}\,,\nn\\
    \alpha^{(3)}_{(1,1)}(\gamma)&: & \gamma&\to\gamma_{\rm Kerr}
    +\Delta^{(2)}_{(0)}+\Delta^{(2)}_{(1)}\,,
\end{align}
and in all other cases we replace $\gamma\to\gamma_{\rm Kerr}$ (including $\Delta A^{\rm 4PN}$).
For example, as $\alpha^{(3)}_{(1,0)}$ appears at 3PM order in $H_{\rm eff}$, we need all corrections up to 2PM order
in order to ensure that our Hamiltonian remains correct up to 5PM.
An exception is the two non-spinning parameters $\alpha^{(2)}_{(0,0)}$ and $\alpha^{(3)}_{(0,0)}$,
wherein we do not include $\Delta^{(3)}_{(0)}$ and $\Delta^{(2)}_{(0)}$, respectively, 
as the non-spinning 5PM contribution is not included from known perturbative PM results.
We instead correct using the 4PN term $\Delta A^{\rm 4PN}$.
The result of this procedure is an effective Hamiltonian that now depends \emph{explicitly} on ($r$,$p_r$,$p_\phi$).
As this procedure changes the nature of the resummation,
our Hamiltonian is therefore different from the one encountered in the \texttt{SEOB-PM} scattering model~\cite{Buonanno:2024vkx}.

A consequence of treating $\gamma$ via Eq.~\eqref{eq:gamma} is
  that it vanishes wherever $H_{\rm Kerr}$ does. For example, in the
  non-spinning limit, this leads to a pole at the Schwarzschild
  horizon $r=2M$ in the $\alpha^{(n)}_{(i,j)}$
  coefficients~\eqref{eq:deltaA}, which contain powers of $1/\gamma$,
  and consequently in the $A$-potential.  This would not necessarily
  occur if $\gamma$ was defined as $\gamma=E_{\rm eff}/\mu$, as the
  effective Hamiltonian does not need to vanish at the horizon.  Since
  the pole appears at a very small separation, it does not prevent
  evolving the binary until the merger time (which occurs before the time 
of horizon crossing) and obtaining an accurate
  inspiral-merger-ringdown waveform.  However, the presence of the pole
  impacts the shape of the $A$-potential (and of the overall Hamiltonian) in the strong-field regime
  and might limit the flexibility that can be gained from incorporating
  higher-order calibration parameters in the EOB Hamiltonian. For
  example, adding a 5PN calibration parameter $a_6 u^6$ to the
  $A$-potential via Eq.~\eqref{eq:PNcorrection}, similar to the one
  used in \texttt{SEOBNRv5}, does not improve the agreement of the non-spinning 
  \texttt{SEOBNR-PM} against NR simulations as effectively as the
  corresponding parameter in \texttt{SEOBNRv5}. 
  We recall that \texttt{SEOBNRv5} is highly faithful to NR simulations in the 
  non-spinning limit, with a median mismatch of $4.6 \times 10^{-5}$ and only $2$ 
  cases having a mismatch $ \gtrsim 1 \times 10^{-4}$, out of 40 SXS NR waveforms.
  In the future, we
  will explore different strategies to interpret $\gamma$, and seek for a 
more suitable choice of the NR calibration parameters tailored to the 
particular structure of the PM terms, which can enable
  a more effective NR calibration of the spinning model.

  \begin{figure*}[t!]
    \includegraphics[width=.46\textwidth]{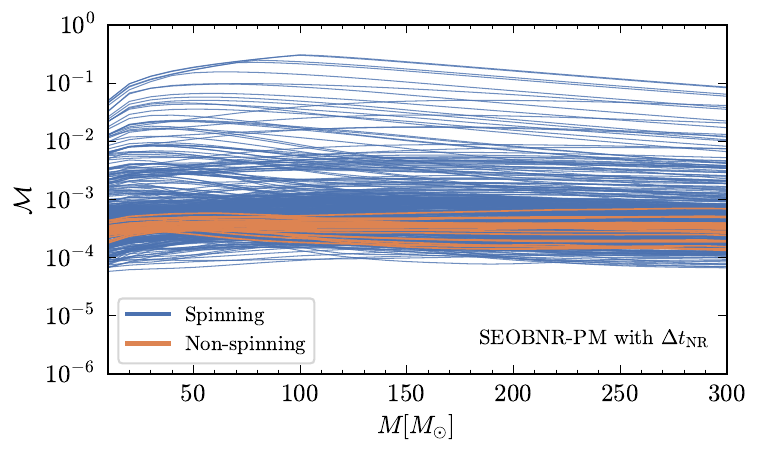}
    \includegraphics[width=.46\textwidth]{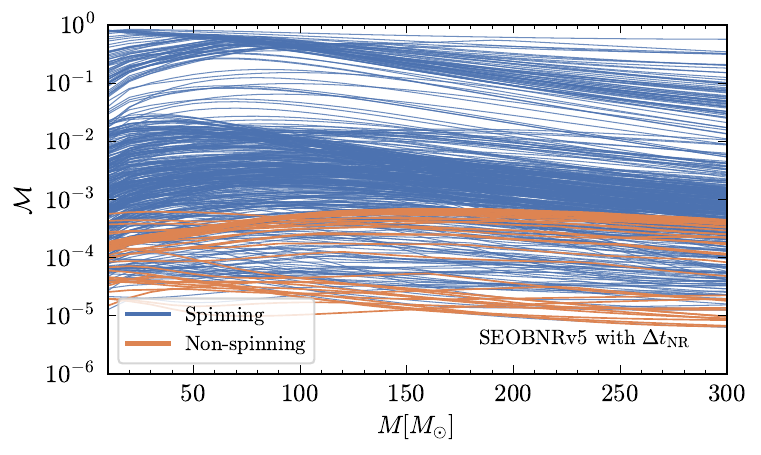}
    \captionsetup{skip=5pt}
    \caption{
      Mismatch over the binary's total-mass range $10 M_{\odot} \leq M \leq 300 M_{\odot}$ for the \texttt{SEOBNR-PM} and \texttt{SEOBNRv5} models calibrated using only the $\Delta t_{\rm{NR}}$ parameter. The study uses 441 SXS NR waveforms, and focuses on the $(\ell,m)=(2,2)$ mode. Spinning and non-spinning configurations are shown in blue and orange respectively.
    }
    \label{fig:mismatch_spaghetti}
  \end{figure*}
  
\section{Fits of the NR-calibration parameters }

Since NR simulations are only available at discrete parameter values, to generate the model for generic configurations we need to fit the NR calibration parameter $\Delta t_{\mathrm{NR}}$ across the ($q, \chi_1, \chi_2$), or equivalently ($\nu, \chi_+, \chi_-$), parameter space. We do this hierarchically, fitting first non-spinning and then aligned-spin configurations, using the same ansatz as in \texttt{SEOBNRv5}. 
The final expression for \texttt{SEOBNR-PM} reads
\begin{align}
    \Delta t_{\mathrm{NR}} &= \nu^{-1/5 + 12.73 \nu} \big(113115.96 \nu^{3} - 25626.22 \nu^{2}\\
    &\quad - 1457.38 \nu - 60.17\big) \nn\\
    &+ \nu^{-1/5} \big(195.45 \nu^{2} \chi_{+} - 190.15 \nu^{2} \chi_{-} + 52.14 \nu \chi_{+}^{2}\nn\\
    &\quad- 72.80 \nu \chi_{+} \chi_{-} - 72.56 \nu \chi_{+} + 50.90 \nu \chi_{-}^{2} \nn\\
    &\quad- 15.76 \nu \chi_{-} + 0.24 \chi_{+}^{4} - 1.51 \chi_{+}^{3} + 4.23 \chi_{+}^{2} \chi_{-} \nn \\
    &\quad- 10.51 \chi_{+}^{2} + 1.22 \chi_{+} \chi_{-}^{2} + 18.89 \chi_{+} \chi_{-}\nn\\ 
    &\quad+ 10.10 \chi_{+}- 10.50 \chi_{-}^{2} + 17.08 \chi_{-}\big)\,. \nn
\end{align}
For completeness, we also report the fit used in the variation of the \texttt{SEOBNRv5} model shown in Fig.~\ref{fig:mismatch}, in which only this parameter is tuned to NR:
\begin{align}
    \Delta t_{\mathrm{NR}} &= \nu^{-1/5 + 4.97 \nu} \big(- 4091.55 \nu^{3} + 2493.06 \nu^{2}\\
    &\quad - 205.61 \nu - 53.99 \big)\nn \\
    &+ \nu^{-1/5} \big( 45.32 \nu^{2} \chi_{+} - 874.81 \nu^{2} \chi_{-} + 71.25 \nu \chi_{+}^{2}\nn\\
    &\quad+ 65.96 \nu \chi_{+} \chi_{-} + 3.72 \nu \chi_{+} - 100.55 \nu \chi_{-}^{2} \nn\\
    &\quad+ 160.60 \nu \chi_{-} - 89.85 \chi_{+}^{4} - 43.49 \chi_{+}^{3}- 1.07 \chi_{+}^{2} \chi_{-} \nn\\
    &\quad+ 59.49 \chi_{+}^{2} + 0.94 \chi_{+} \chi_{-}^{2} - 17.51 \chi_{+} \chi_{-} \nn\\
    &\quad+ 36.27 \chi_{+} + 26.08 \chi_{-}^{2} + 15.37 \chi_{-} \big)\,. \nn
\end{align}
The complete \texttt{SEOBNRv5} model also contains calibrations of the $a_6$ and $d_{\rm SO}$ parameters, together with a different fit for $\Delta t_{\mathrm{NR}}$~\cite{Pompili:2023tna}.

\section{Mismatch in terms of the total mass}

In Fig.~\ref{fig:mismatch_spaghetti} we show the mismatch over the binary's total mass range $10 M_{\odot} \leq M \leq 300 M_{\odot}$ for the \texttt{SEOBNR-PM} and \texttt{SEOBNRv5} models calibrated using only the $\Delta t_{\rm{NR}}$ parameter, against the 441 SXS NR waveforms used in this work. Spinning and non-spinning configurations are shown in blue and orange, respectively. This provides a complementary picture to Fig.~\ref{fig:mismatch}, which only presents the maximum mismatch across the total mass range for each configuration.

\newpage

\bibliography{./../wqft_spin}

\end{document}